\documentclass[twocolumn,aps]{revtex4}
\usepackage{graphicx}
 
\begin{document}

\title{Stochastic Modeling of Single Molecule Michaelis Menten Kinetics}
\author{ Mahashweta Basu} \email{mahashweta.basu@saha.ac.in}
\author{P. K. Mohanty } \email{pk.mohanty@saha.ac.in}
\affiliation{ TCMP Division, Saha Institute of Nuclear Physics,1/AF Bidhan Nagar, Kolkata 700064, India}
\begin{abstract}  
 We develop an general formalism of single enzyme  kinetics in two  dimension   
where substrates diffuse stochastically  on  a square lattice in presence of
disorder. The dynamics of the model could be decoupled
effectively to {\it two} stochastic processes, (a) the substrate arrives at
the enzyme site  in intervals which fluctuates in time and (b) the enzymatic
reaction takes place at that site stochastically.
 We argue that  distribution of arrival time is a two parameter function
specified by  the  substrate and the disorder densities,  and that it
correctly reproduce the distribution of turnover time obtained from
Monte-Carlo simulations of single enzyme kinetics in two dimension, both
in absence  and presence of disorder. The decoupled dynamics
model is simple to implement and generic enough to describe both normal and
anomalous diffusion of substrates. It also suggests that  the diffusion 
of substrates in the single enzyme systems could  explain the different distributions of 
turnover time observed in recent experiments.
\end{abstract} 
\maketitle
 \section{Introduction}
Biology of life solely depend on  complex 
chemical reactions brought about by certain specific enzymes.
 Enzymes are known to facilitate  reaction rates without taking  part in 
reactions directly\cite{book}.  As catalysts, 
enzymes  are receiving  increasing importance  from  other branches 
of science. The mechanism of enzymatic activity  is  one of the most 
fascinating field  of  physical chemistry. 
 In general in a chemical reactions like ($A + B \rightarrow P$) the rate of reaction 
depends linearly on the initial concentration of the reactants. 
But  in several enzymatic reactions, where 
enzyme (or catalyst)  $E$ and substrate (or reactant) $S$ form a 
product $P$,  the rate of reaction is known to saturate \cite{b} with increase of substrate 
concentration $[S]$ (although  the linearity remain valid for low concentration $[S]$), 
which was explained by Leonor Michaelis and Maud Menten \cite{Michaelis} in $1913$. 
In their formulation,  it was postulated that enzyme (or catalyst)  $E$ and substrate 
(or reactant) $S$ form a product $P$ through a intermediate  enzyme-substrate complex $ES$ as
\begin{equation}
 {\rm E+S} \mathop{\rightleftharpoons}^{k_{1}}_{k_{2}} {\rm ES}\mathop{\rightarrow}^{k_{3}} E+P
\label{eq:main}
\end{equation}
 Using law of mass action it was shown that the reaction  proceeds with velocity 
${{V [S]}\over{([S] +K)}}$, where $V$ and $K$ are constants which depend on the 
rates of the reaction.
Over years this formulation,  popularly known as  Michaelis-Menten  (MM) kinetics, has  
found applications in different enzymatic reactions\cite{b} and in several other 
systems \cite{myxo,drug}.
Although Michaelis-Menten kinetic model  is based on law of mass action  its extension 
for ordinary or fractal spacial dimesions \cite{Sava}
with or without  disorder have been attempted .  To incorporate effect of diffusion and noise 
these multi-enzyme systems  have been studied  in two dimension using  Monte-Carlo 
simulation\cite{Berry}. 

Recent development of experimental techniques allow one to do single molecule experiments \cite{Lu, Xie}, where 
substrates react stochastically with a single  enzyme held at a fixed position. Thus, formation of the 
products here form a  discrete time series, with interspike intervals $\tau$ distributed  as $f(\tau)$. 
It has been shown that the rate of reaction  $\langle \tau\rangle^{-1}$ follows 
MM law  independent of the functional form of $f(\tau)$. Thus the distribution of 
turnover time $f(\tau)$, rather than  MM law,  could be taken as the characteristic property 
of any particular single enzymatic reaction system. To understand the  variations 
of $f(\tau)$ observed in experiments several theoretical models \cite{Kou, models} 
have been proposed.  In Ref. \cite{Kou},  density of  $S$, $P$, $E$ and $ES$ are replaced 
by  their corresponding probabilities to explain the form of $f(\tau)$. However 
effect of diffusion  and noise, which are known to modify the theoretical description 
in other low dimensional systems\cite{lowD} have not been incorporated in these studies.

In this article we study  the single enzyme system  in two dimension using  Monte-Carlo simulation and 
calculate $f(\tau)$. The substrate molecules $S$, here, 
are allowed to diffuse  on the lattice and  react  stochastically with the enzyme $E$ fixed at a 
particular site. We also propose an alternative simple method which illustrate that  the  dynamics of this two dimensional  system which explicitly 
takes care of the diffusion of the substrates can be decoupled 
into two separate dynamics; 
(a) the substrates arrive at the \textit{enzyme site} in intervals 
of $\zeta$ which varies stochastically and its 
distribution is $P(\zeta)$, and (b) the  enzymatic reaction (\ref{eq:main}) 
takes place at the  \textit{enzyme site}. 
We further argue that $P(\zeta)$  is a 
universal two parameter function  independent of whether 
the system has  quenched or annealed disorder, 
or if the diffusion of substrates is normal or anomalous.
Given a  $P(\zeta)$, the decoupled dynamics can be 
implemented using a single-site  stochastic simulation, which greatly 
simplify the study  by overlooking the full dimensionality of the  system.

The article is organized as follows. In section \textbf{II} we describe the Monte-Carlo 
simulation of single enzyme kinetics in two dimension  and calculate the distribution
of turnover time $f(\tau)$.  Description of the decoupled dynamics,  
characterization  of distribution of substrate arrival time  $P(\zeta)$, and 
comparison of   $f(\tau)$  obtained from the decoupled dynamics model with that obtained 
from  Monte-Carlo simulation are 
discussed in section \textbf{III}. In section \textbf{IV}, we  show that  the decoupled dynamics 
model correctly reproduce  $f(\tau)$ both in absence and presence of disorder. Finally, 
in section\textbf{V} we conclude the study with some discussions.

\section{Monte-Carlo simulation of single enzyme kinetics in two dimension}

To  model the single enzyme kinetics  in spacial dimension $d=2$ (SEK2d) we  take 
a $(L\times L)$ square lattice with periodic boundary condition 
where sites are labeled by two positive 
integers $(i,j)= 1,2,\dots L$. The enzyme is placed at the site 
$i=L/2=j$, which is denoted as {\it enzyme-site}.  
At very other site,  we have a site variable 
$s_{ij}=0,1,2$ which corresponds to the fact that a site is either 
vacant, occupied by a substrate $S$ or by a product $P$  respectively. 
The  restriction that each site can be occupied by 
at most one particle captures the hardcore
interactions present  among the molecules.
The density of the substrates $[S]$ is denoted by $\rho = \frac{N_S}{L^2}$
where $N_S$ is the number of substrates.

The dynamics of the system can be described as follows.
Far from the {\it enzyme-site} the substrates and the products
simply diffuse by making an unbiased move to one of the neighbouring  vacant site.
 These moves are not allowed if all the neighbours are occupied.
The enzyme, which is pinned to the {\it enzyme-site}, can either be in state 
$E$ or makes a enzyme substrate complex $ES$.
The  enzymatic reaction takes place as follows.
If the enzyme is in state $E$ and one of the neighbour is occupied by $S$, 
then $E+S\to ES$  by making the neighbouring site  vacant. This process occurs 
with probability $k_1$. Otherwise when $ES$ is present at the {\it enzyme-site} and 
one of the  neighbour is vacant,  then either reaction $ES\to E+S$  or $ES\to E+P$
occurs  with rates $k_2$ and $k_3$ respectively. The substrate $S$ or the product $P$ which is
yield in these reactions goes to the vacant site. Note that, the system is updated  in 
discrete time, hence the probability  $k$ that  an reaction occurs  is same as 
$r dt$, where $r$ is the rate of  that chemical reaction and $dt= 1/L^2$. 

\begin{figure}[hbt]
 \includegraphics[width=7cm]{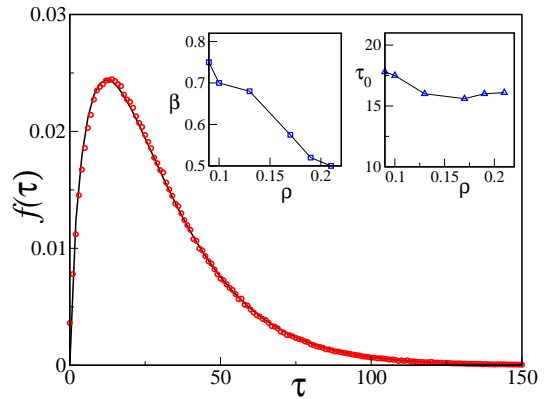}
\caption{ The distribution of turnover time $f(\tau)$ calculated for $\rho = 0.1$ is
 a $\Gamma$-function \ref{eq:gamma} with $\beta =0.7 ,\tau_0=17.5$.Similar fit could be done
for different $\rho$ by taking $\beta$ and  $\tau_0$ as parameters.
The inset shows variation of $\beta$ and $\tau_0$  with $\rho$. Other parameters are
$k_1=1$, $k2=0.001$, $k3=0.03,$ and $L= 100$. 
}
\label{fig:fT} 
\end{figure}

   To calculate the rate of reaction in a multi-enzyme system one
counts the number of products $N_P$ in time and then  $\frac{dN_P}{dt}=  
-\frac{dN_S}{dt}$ gives the rate of reaction. Dependence of  the rate on  
initial substrate concentration $\rho$ is described by  the MM law in  a
regime $\rho\ll1$. In the single enzyme reaction, however, product 
forms in discrete intervals $\tau$  whose distribution is defined  as 
$f(\tau)$. To calculate $f(\tau)$ we have done Monte-Carlo simulations of
SEK2d and  obtain the time interval $\tau$ between formation of 
any two consecutive products. $\tau$ is measured in units of Monte-Carlo sweep (MCS).  
As shown in Fig. \ref{fig:fT}  the distribution of  $\tau$ is  found to be  
a $\Gamma$-function,  
\begin{equation}
f(\tau) = A \tau^\beta \exp(-\tau/\tau_0),
\label{eq:gamma}
\end{equation}  where  both 
the exponent $\beta$ and the cut-off scale $\tau_0$  depend on  density of 
the substrates $\rho$. The insets  of  Fig. \ref{fig:fT} shows  these variations. 
 Note that the distribution $f(\tau)$,  shown in Fig. \ref{fig:fT}, 
differs substantially from  the experimental studies, mainly because, the hardcore 
interaction here allow only four substrates in the neighbourhood of the 
enzyme which decreases the turnover probability $f(\tau)$ for small $\tau$. 
Another possible reason is that the sensitivity of the detector is insufficient in experiments
to  resolve small turnover time and  the observed $f(\tau)$ could be the 
result of an integrated effect.

\begin{figure}
 \includegraphics[width=7cm]{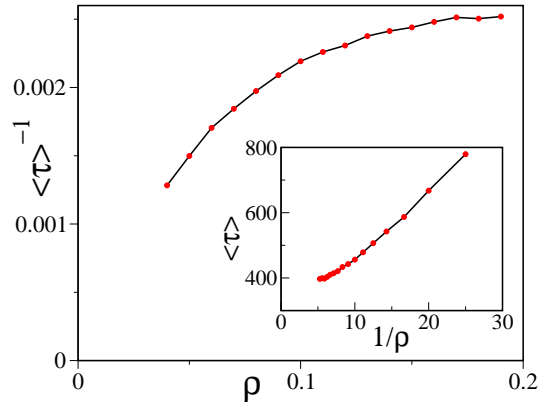}
\caption{The rate $\langle \tau\rangle^{-1}$  obey Michaelis Menten law. 
Inset shows that the inverse rate depends linearly on $\rho^{-1}$ for low densities.}
\label{fig:mmlaw}
\end{figure}

Since in single 
enzyme systems the  average time interval between formation of two consecutive products  
is $\langle\tau\rangle$, the rate of reaction is $\langle\tau\rangle^{-1}$. We have 
calculated $\langle\tau\rangle^{-1}$  for different  $\rho$  and plotted it in 
Fig. \ref{fig:mmlaw}. According to MM law inverse rate of turn over depends linearly
on the inverse rate of substrate density, which is described in inset of Fig. \ref{fig:mmlaw}.
The linear relation holds good for low densities as expected. Thus, 
it is evident that this single enzyme system 
follow  MM law independent of the functional form of $f(\tau)$. This has been 
seen  in  the experiments of single enzyme systems\cite{Xie} recently. 

However, since the distribution  of $\tau$ contains more information than the rate of the reaction
$\langle \tau\rangle$ ,  in this article we give prime 
importance to  $f(\tau)$  and  try to achieve a unified formulation which provide a 
simple method of calculating $f(\tau)$, without doing  a  two dimensional 
simulation. This is done in the next section.

\section{The decoupled dynamics and its stochastic modeling}
 In the  single enzyme systems  two different dynamics act  separately. First, 
that the substrates diffuse (with hard core restrictions) through  the two 
dimensional lattice and arrive stochastically at the  enzyme site. Second, 
that  the enzymatic reaction (\ref{eq:main}) takes place  there. Based on  these decoupled dynamics 
we introduce a stochastic model of single enzyme kinetics as follows. 

\begin{figure}[htb]
 \centering
 \includegraphics[width=5cm, bb=0 450 490 740]{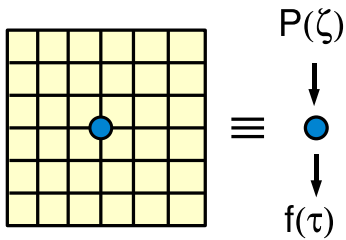}
 \caption{Schematic representation of decoupled dynamics model.}
 \label{fig:cartoon}
\end{figure}

In this model, the  substrate arrives stochastically  at the  enzyme  
site in intervals of $\zeta$ and then the  enzymatic reaction takes 
place  there  according to  (\ref{eq:main}).  
For a given distribution $P(\zeta)$, one can simulate this single 
site dynamics as follows.  The enzyme site has two possible  states, $E$ and 
$ES$. If it is in state $E$ it continues to be in the same sate  until
a substrate $S$ arrives there, and after that it is converted to $ES$ instantly.
Otherwise if the site is in state $ES$  it is converted to $E$ stochastically
either by forming a  product $P$ with rate $k_3$ or by breaking the complex 
$ES\to E+S$ with rate $k_2$.  However, in the mean time if the substrate arrives 
while the site is in state $ES$ it falls off. Figure \ref{fig:cartoon} describes  
this schematically.

\begin{figure}[htb]
 \centering
 \includegraphics[width=6cm]{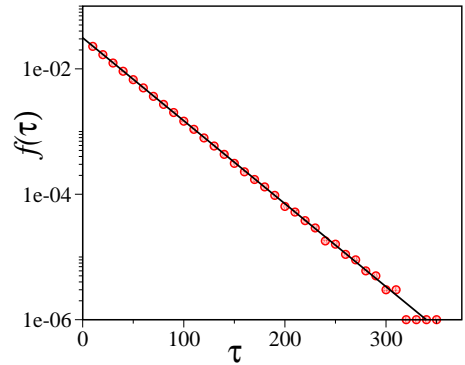}
 \caption{ Exact result (solid-line) of the turnover distribution $f(\tau)=A (1-k_3)^\tau$ for  
$P(\zeta) = \delta (\zeta-1)$ is compared with the corresponding simulation of decoupled 
dynamics(symbols). Here, $ k_1=1, k_2=0.001, k_3=0.03$.}
 \label{fig:exp}
\end{figure}

Thus,  given $P(\zeta)$  one can  calculate  $f(\tau)$ numerically  using 
the above stochastic  model  by noting down the time difference 
between formation of any two consecutive products. For the simplest  
 model  system,  where  substrates are always available at the 
\textit{enzyme site}, $f(\tau)$ can be calculated analytically. In this case,
 $P(\zeta) = \delta (\zeta-\Delta)$, 
where $\Delta$ is the unit of time in discrete-time-update (here  
$\Delta=1~ MCS$). Thus, the \textit{enzyme site} is always in the state $ES$  and 
form products stochastically with with rate $k_3$. Of course, $ES$ breaks up 
with rate $k_2$ to from $E+S$,  but then instantly  $E\to ES$ as  other 
substrates are available at the \textit{enzyme site}. As expected, this results in an exponential distribution 
$f(\tau)= A (1-k_3)^\tau$, with $A$ being the normalization constant. The exact 
form of  $f(\tau)$ is compared with that obtained from numerical simulation 
of decoupled dynamics in Fig. \ref{fig:exp}.  

In the next section we calculate  $P(\zeta)$ for the single enzyme system. 

\subsection{  Distribution of arrival time $P(\zeta)$} 
  First let us consider the system without hardcore interaction among the substrates, 
which in turn allows each  site of the lattice  to  accommodate more than one substrate. 
Hence each substrate in the system  does a simple random walk in two dimension. 
Starting from  the \textit{enzyme site} $E$, a  simple random walk would end at a distance within 
 $r$ and $r+dr$ in time $\zeta$ with probability 
\begin{equation}
 Q(\zeta)=\frac{1}{\sqrt{\pi \zeta} } e^{-r^2/\zeta} dr.
\label{eq:rwalk}
\end{equation}
Thus, a substrate residing in the region $r$ and $r+dr$ would reach the enzyme
$E$ with   the same probability $Q(\zeta)$. Therefore  the arrival probability of 
a substrate (from anywhere within the lattice) at  $E$ is  
\begin{equation}
 P(\zeta)= \frac{1}{\sqrt{\pi \zeta} }\int e^{-r^2/\zeta} dr \sim C~{\rm erf}({A\over \sqrt{\zeta}}), 
\label{eq:zeta1}
\end{equation}
where $A$ depends on the upper limit of the integral and $C$ is a normalization constant. 
Note that, for a finite system  $\zeta$ can not be larger than a typical  time scale 
$\zeta_0$, which  introduces a cut-off  $\zeta_0$ to $P(\zeta)$ , 
\begin{equation}
 P(\zeta)=C~{\rm erf}({A\over \sqrt \zeta})\exp(-{\zeta\over\zeta_{0}}),
\label{eq:zeta2}
\end{equation}

Interaction among substrates, e.g.  hardcore interaction in  our study, further 
modify Eq. (\ref{eq:zeta2}) in two ways. It  decrease
the cutoff $\zeta_0$ which now depends strongly on the density of particles $\rho$,
and it also modifies the exponet of $\zeta$  which appears in the argument of erf 
function; 
\begin{equation}
 P(\zeta)=C~ {\rm erf}({A\over \zeta^\alpha})\exp(-{\zeta\over\zeta_{0}}).
\label{eq:pzeta}
\end{equation}
In fact $\alpha$ depends on the  exponent  $\gamma$ which appears  
in the mean square displacement of a tagged  particle as 
$$\langle r^2\rangle \sim t^\gamma.$$  For non-interacting systems  
$\gamma=1$, resulting in $\alpha=\gamma/2=1/2$, which is 
consistant with  Eq. (\ref{eq:zeta1}).
In presence of hardcore interaction, however,  $\langle r^2\rangle$  gets  modified 
by  a logarithmic  term \cite{Henk} in the asymptotic regime. 
Since in the enzymatic systems we are rather interested in  the small time limit, 
the logarithmic  term  appears as an 
effective exponent $\gamma^\prime > 1$ which results in $\alpha>1/2$. 

\begin{figure}[hbt]
 \centering
 \includegraphics[width=7cm]{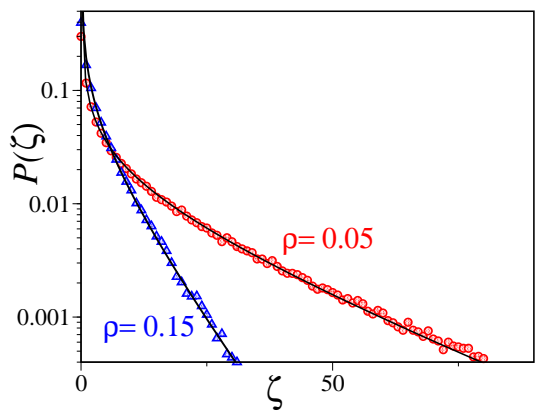}
 \caption{  Arrival time distribution $P(\zeta)$ plotted for two different densities 
 $\rho=0.05,0.15$ with $L=100$. The solid lines here corresponds to the best fit 
according to Eq. (\ref{eq:pzeta}) with $\alpha=0.64,.7$ ,  $\zeta_0=27.5,7.8$ 
and  $C=1.05,2$ for respective densities. For both the densities $A=0.1$ }
 \label{fig:p_zeta}
\end{figure}

    In the following we  calculate $P(\zeta)$ numerically.  
Substrates, initially placed randomly, are allowed to 
diffuse  on a two dimensional square lattice (with periodic boundary condition).  
Arrival time of any of the substrates at a pre-decided fixed site is recorded. 
The  distribution $P(\zeta)$  is calculated from the difference  between  
any two consecutive arrival times.   
The numerical data of $P(\zeta)$ obtained  
for two different density of substrates $\rho=0.05,0.15$  could be fitted  
to Eq. (\ref{eq:pzeta}) by using $\alpha$ and $\zeta_0$ as fitting parameters
(see Fig. \ref{fig:p_zeta}). An excellent  fit with the numerical data  strongly supports  
the theoretical form of $P(\zeta)$ described by (\ref{eq:pzeta}).  

\subsection{ Distribution of turnover time  $f(\tau)$ for single enzyme system}
\begin{figure}[hbt]
 \centering
 \includegraphics[width=7cm]{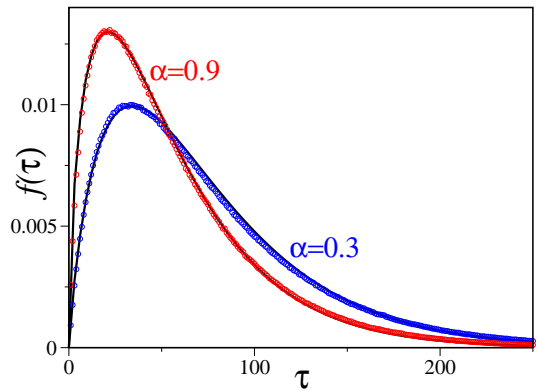}
 \caption{$f(\tau)$ obtained from the decoupled dynamics model is 
a  $\Gamma$-function with $(\beta,\tau_0) = (.85,40), (0.61,34.5)$ for $(\alpha,\zeta_0)=(0.3,50), (0.9,50)$ respectively.}
 \label{fig:ftau}
\end{figure}

  We have already  seen that  $f(\tau)$ can be obtained  numerically 
for any given $P(\zeta)$ using the  decoupled dynamics model discussed 
earlier in this section. For the single enzyme problem $P(\zeta)$ is 
generically described by (\ref{eq:pzeta}), a function with \textit{two} parameters 
$\alpha$ and $\zeta_0$ (unless otherwise specified 
we take $A=0.1$  all through the article). For the same system we know from the 
Monte-Carlo simulations of single enzyme systems
that $f(\tau)$  is a $\Gamma$-function. First let us check 
that if $P(\zeta)$ given by (\ref{eq:pzeta})  
truly produce  the same $\Gamma$-function  in the  decoupled dynamics model.
Figure \ref{fig:ftau} demonstrates that numerical data for $f(\tau)$ obtained 
from the  decoupled dynamics model fits well with  a $\Gamma$-function.  
 
\begin{figure}[hbt]
 \centering
 \includegraphics[width=7cm]{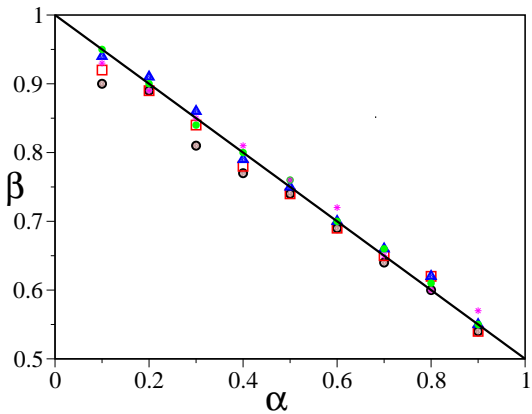}
 \caption{The figure shows $\beta$ versus $\alpha$, calculated form the decoupled 
dynamics model, for $\zeta_0=15,25,50,70$ and $100$ (different symbols). 
The relation $\beta=1-\alpha/2$ (solid-line) is satisfied for all values of $\zeta_0$.  }
 \label{fig:ab}
\end{figure}

\begin{figure}[hbt]
   \centering
   \includegraphics[width=7cm]{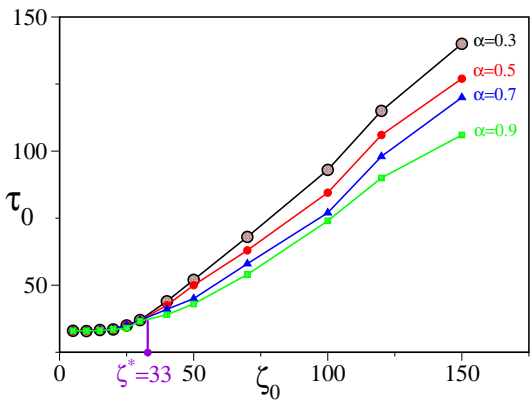}
   \caption{The figure shows $\tau_0$ versus $\zeta_0$, calculated form the decoupled dynamics 
model, for different $\alpha$. It is evident that $\tau_0$ is independent 
of $\alpha$  for $\zeta\le \zeta^*$.}
   \label{fig:zeta_tau0}
\end{figure}

Hence from the decoupled dynamics we infer that the parameters of $f(\tau)$ depends on  
$P(\zeta)$. Thus, for the complete description of the model it is reasonable enough 
to characterize $(\beta, \tau_0)$ in terms of $(\alpha,\zeta_{0})$ which is done 
in the following.  Figure  \ref{fig:ab} shows the dependence of $\beta$ on $\alpha$ for 
different values of $\zeta_0=15,25,50,70,100$.  Clearly from the numerical data  it 
emerges that 
\begin{equation}
\beta= 1-\frac \alpha 2. 
\label{eq:ba}
\end{equation}
 Similarly Fig. \ref{fig:zeta_tau0} describes the dependence of $\tau_0$ on 
$\zeta_0$ for different values of $\alpha=0.3, 0.5, 0.7, 0.9$. From the figure it it  is clear 
that $\tau_0$ is independent of both $\alpha$ and $\zeta_0$ below a specific 
time scale $\zeta^*$. For $\zeta>\zeta^*$, however, functional dependence of 
$\tau_0$ on $\zeta_0$ varies with choice of $\alpha$. We  notice that the 
time scale $\zeta^*$ originates from  the product formation rate  
$k_3$ (and do not depend on other rates, like $k_1$ or $k_2$) as 
$$\zeta^*  \simeq \frac 1 k_3.$$  

 Thus,  to compare $f(\tau)$  obtained from the two dimensional simulation of single enzyme 
kinetics with that of the decoupled dynamics model it is convenient to restrict $\zeta_0$ in the 
regime $\zeta_0\le \zeta^*=1/k_3$.  In fact, without loss of generality one may take 
\begin{equation}
\zeta_0=\zeta^*=1/k_3.
\label{eq:z0k3}
\end{equation}
Otherwise, when $\zeta_0> \zeta^*=1/k_3$, the choice of $\tau_0$ would depend 
explicitly on the value of $\alpha$ (see Fig. \ref{fig:zeta_tau0}). A compatible choice, 
which we use here, is $\zeta_0 =\tau_0$.  Thus, to obtain  the return time distribution   
 $f(\tau)$  which is characterized by two parameters $\beta$ and $\tau_0$  we  make a choice 
of $P(\zeta)$ with parameters 
\begin{eqnarray}
 \alpha= 2(1-\beta) &;&   \zeta_0=\tau_0\cr
{\rm and } ~~k_3&=& 1/\tau_0\label{eq:k3}
 \end{eqnarray}
in the stochastic simulation of decoupled dynamics, where the last line of (\ref{eq:k3}) corresponds to  
the fact that $k_3$ is constrained by (\ref{eq:z0k3}), and we have used  $\zeta_0 =\tau_0$.
In the next section we will check if  choice (\ref{eq:k3}) in decoupled dynamics model
reproduce the $f(\tau)$ same as that of SEK2d.
 
\subsection{Comparison of decoupled dynamics model with Monte-Carlo simulation}
From the last section  we have seen that the stochastic decoupled dynamics model
generates a correct form of $f(\tau)$  for any  $P(\zeta)$  given by  
Eq. (\ref{eq:pzeta}). To make a correspondence with the SEK2d one needs a
 mapping between two parameters  of $f(\tau)$, namely ($\beta,\tau_0$), with 
($\alpha,\zeta_0$) of $P(\zeta)$. From the numerical characterization of  $P(\zeta)$
described in Fig. \ref{fig:ab} and \ref{fig:zeta_tau0}, we  have seen that the mapping 
is not unique. The same  values of  ($\beta,\tau_0$) can be obtained from different 
($\alpha,\zeta_0$) by varying the reaction rates of stochastic model. We take  an 
advantage of this situation to make a choice   (\ref{eq:k3}).  Now let us verify  if 
such choice works.

\begin{figure}[hbt]
 \centering 
\includegraphics[width=7cm]{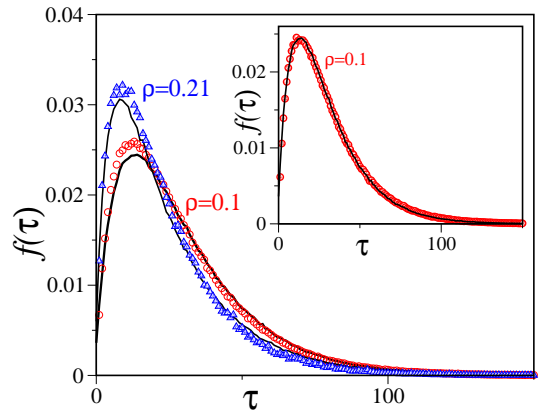}
 \caption{ $f(\tau)$ obtained from decoupled dynamics model is compared with 
 the same obtained from SEK2d for two different $\rho=0.1,0.21$. Other parameters are
$k_1=1, k_2=0.001$.}
 \label{fig:match}
\end{figure}

      In Fig. \ref{fig:match} we have compared $f(\tau)$  obtained from single 
enzyme kinetics in two dimension and   \textit{corresponding} decoupled dynamics model for 
two different densities $\rho = 0.1, 0.21$. To calculate $f(\tau)$ for 
density $\rho=0.1$, which is a $\Gamma$-function with $\beta=0.70$ and $\tau_0=17.5$, we  
simulate the decoupled dynamics for this system with $\alpha= .6, \zeta_0=17.5$ and 
$k_3=0.057$ as   prescribed  by Eq. (\ref{eq:k3}). Similar procedure is used for $\rho=0.21$ 
where  $\beta=0.5$  and $\tau_0=16.1$. 

  From  Fig. \ref{fig:match} it is clear that  $f(\tau)$  obtained from SEK2d matches 
reasonably well 
with that obtained from corresponding decoupled dynamics. A small discrepancy which  is 
visible for both the densities is possibly due to the fact that  a discretized form of 
$P(\zeta)$ and a cut-off in $\zeta $ has been used in the  numerical simulation for convenience. 
A minor modification of $\zeta_0$  can  compensate for this discrepancy. $f(\tau)$ with 
a modified $\zeta_0=18.5$ for density $\rho=0.1$ is presented in the inset.

From  the above analysis it is evident that the formulation of decoupled dynamics model 
works well for  single enzyme systems  without disorder. In the next section we will 
apply the same formalism to systems with mobile or immobile impurities.

\section{Single enzyme systems in presence of disorder}
In this section we  discuss  the  effect of disorder in SEK2d.  
Disorder can be 
introduced in the single enzyme system  by introducing $N_D$  impurities in  
the two dimensional lattice. In case of annealed disorder, the impurities  are 
allowed to diffuse in the system. Thus, it is equivalent to a late time   single enzyme 
system without disorder when products are considered as diffusing impurities. Compared to 
systems without disorder, here, only the cut-off time scale 
$\tau_0$ is larger. So, in  this section  we choose to work in the 
system of quenched disorder in details.
The impurities in these systems are immobile and do not diffuse through the system. 
Thus the substrates are restricted  from visiting these impurity sites, and the effective  
 dimension of these system are $d<2$. Simple diffusion on these fractal 
lattices \cite{fractal} are anomalous where mean square displacement  is given by 
$ \langle r^2\rangle  \sim t^\gamma$ with $\gamma<1$. 
Note that $\gamma=1$ corresponds to the normal diffusion. 
Obviously,  $\gamma$ depends on the disorder density $\rho_D= N_D/L^2$ for small 
disorder densities. Beyond a critical density $\rho_D> 0.4$ \cite{Sahimi} 
however the disorder sites  percolate and the substrates remain confined in 
certain islands.

\begin{figure}[htb]
 \includegraphics[width=6cm]{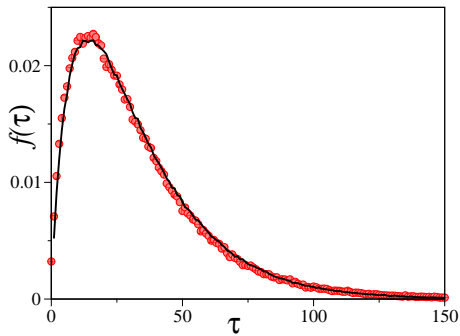}
\caption{$f(\tau)$ obtained from decoupled dynamics model (solid line) 
is compared with that of SEK2d (symbols) for disorder density $\rho_D=0.1$ and 
density of substrates $\rho=0.1$. Here $k_3=0.03$ for SEK2d and $k_3={1\over\tau_0}$
for decoupled dynamics model.Other parameters are $k_1=1, k_2=0.001$.}
\label{fig:fT1}
\end{figure}

The anomalous diffusion is expected to change the behavior of arrival time 
distribution $P(\zeta)$. In particular, if we ignore hardcore interaction among 
the substrates,  Eq. (\ref{eq:zeta2}) gets modified  in presence of disorder as
\begin{equation}
P(\zeta)=C~{\rm erf}({A\over \zeta^{\gamma/2}})\exp(-{\zeta\over\zeta_{0}}).
\label{eq:zeta3}
\end{equation}
Again, the hardcore interaction  introduces a   a logarithmic correction term to 
the  mean square displacement, as we have seen in systems without disorder in section {\bf III} A. 
This correction may appear in Eq. (\ref{eq:zeta2})  as an effective exponent 
$\alpha\ne \gamma/2$. Next, we ask if  $f(\tau)$  is still a $\Gamma$-function in presence of 
quenched disorder, and  if the decoupled 
dynamics with  a choice (\ref{eq:k3}) reproduces correct $f(\tau)$. It turns out that 
answer to both the questions are positive  which is described in Fig. \ref{fig:fT1}. 
To calculate $f(\tau)$  we did a Monte-Carlo simulation of single 
enzyme kinetics on a two dimensional square lattice  ($100\times100$) with 
$\rho_D=0.1$ and $\rho=0.1$. 
This  gives rise to $f(\tau)$ as given in (\ref{eq:gamma}) 
with $\beta=0.72$ and $\tau_0=20$. From Eq. (\ref{eq:k3}) one 
would then expect that $\alpha =0.57$ and $\zeta_0=20$ would generate the same 
$f(\tau)$ from  decoupled dynamics model if $k_3=0.05$.
Calculated  $f(\tau)$ for both the  decoupled dynamics model(solid line)
 and SEK2d (symbols) are  shown in Fig. \ref{fig:fT1}.
Clearly  $f(\tau)$ for the decoupled dynamics model compares well with SEK2d.
 Note that, as expected,  the value 
of $\alpha=0.57$ in case of disorder is smaller than $\alpha=0.60$   
for the system without disorder.

A very good agreement of decoupled dynamics  with that of the  SEK2d
even in presence of quenched disorder suggests that the former 
stochastic model provide an alternative simple  description of single enzyme systems.

\section{Conclusion}

 In conclusion  we have studied the Monte-Carlo dynamics of  single enzyme system
in two dimension both in presence and absence  of quenched disorder. In both cases
the  distributions $f(\tau)$ of turnover time $\tau$ are found to be a
$\Gamma$-function described by the exponent $\beta$ and the cutoff scale $\tau_0$. We
argue that the dynamics of the single enzyme system could be decoupled to two
stochastic processes; first that the substrates arrive at the \textit{enzyme site} in intervals
which fluctuate in time, and second that the reaction occurs at the enzyme site. 
We argue that the distribution of substrate arrival time $P(\zeta)$ is a specific 
function (\ref{eq:pzeta}) of two parameters $\alpha$ and $\zeta_0$. This functional form (\ref{eq:zeta2}) 
is generic  for systems with or without disorder.  By choosing 
these parameters  $\alpha$ and  $\zeta_0$ according to Eq. (\ref{eq:k3}),
we further show that the decoupled dynamics model correctly reproduce  $f(\tau)$ 
obtained from  the Monte-Carlo simulation of single enzyme kinetic models defined on a square lattice.

   Distribution  of turnover time $f(\tau)$, rather than the rate $\langle \tau\rangle^{-1}$, 
 is a characteristic feature of a specific single enzyme  system. Following the experiments in 
these systems\cite{Xie}  several  theoretical models  have been  proposed  to explain  
the underlying cause  behind the variation  in observed $f(\tau)$. In these formulations,
after the  enzyme-substrate complex breaks up by forming a product the enzyme returns 
to its normal state with a {\it delay}. Such a delay  is considered to be essential 
\cite{Kou}, without which \textit{asymptotic decay} of $f(\tau)$ for large $\tau$ can  not 
be obtained. Further, all the  reaction rates  there are assumed to vary 
stochastically, which is attributed to the dynamic disorder associated with 
several  possible conformal variations of enzyme and enzyme-substrate complex.
Alternatively,  here we  show that a simple Michaelis-Menten kinetics (\ref{eq:main})  in presence 
of diffusion and noise could produce the same $f(\tau)$  observed in experiments.

\end{document}